\documentclass[conference,tikz, margin=3mm]{IEEEtran}
\IEEEoverridecommandlockouts
\usepackage{cite}
\usepackage{amsmath,amssymb,amsfonts}
\usepackage{algorithmic}
\usepackage{graphicx}
\usepackage{textcomp}
\usepackage{xcolor}
\usepackage{mathtools}
\def\BibTeX{{\rm B\kern-.05em{\sc i\kern-.025em b}\kern-.08em
    T\kern-.1667em\lower.7ex\hbox{E}\kern-.125emX}}
\usepackage{smartdiagram}
\usepackage{pst-node}
\usepackage{lipsum}
\usepackage{stfloats}

\usetikzlibrary{arrows.meta,
                chains,
                positioning,
                shapes.geometric
                }
\DeclareUnicodeCharacter{202F}{\,}
\begin{document}

\title{Smart and Context-Aware System employing Emotions Recognition\\
%{\footnotesize \textsuperscript{*}Note: Sub-titles are not captured in Xplore and should not be used}
%\thanks{Identify applicable funding agency here. If none, delete this.}
}
\makeatletter
\newcommand{\linebreakand}{%
  \end{@IEEEauthorhalign}
  \hfill\mbox{}\par
  \mbox{}\hfill\begin{@IEEEauthorhalign}
}
\makeatother

\renewcommand\IEEEkeywordsname{Keywords}

\author{\IEEEauthorblockN{Stuti Sehgal}
\IEEEauthorblockA{\small\textit{Department of Computer Science} \\ 
\textit{and Engineering}\\
\textit{SRM Institute of Science and Technology,}\\
Kattankulathur, Tamil Nadu– 603203, \\
India \\
ss3537@srmist.edu.in}
\and
\IEEEauthorblockN{Harsh Sharma}
\IEEEauthorblockA{\small\textit{Department of Computer Science} \\ 
\textit{and Engineering}\\
\textit{SRM Institute of Science and Technology,}\\
Kattankulathur, Tamil Nadu– 603203, \\
India \\
hs7685@srmist.edu.in}
\and
\IEEEauthorblockN{Akshat Anand}
\IEEEauthorblockA{\small\textit{Department of Computer Science} \\ 
\textit{and Engineering}\\
\textit{SRM Institute of Science and Technology,}\\
Kattankulathur, Tamil Nadu– 603203, \\
India \\
aa4407@srmist.edu.in}
}

\maketitle

\begin{abstract}
People have the ability to make sensible assumptions about other people's emotional states by being sympathetic, and because of our common sense of knowledge and the ability to think visually. Over the years, much research has been done on providing machines with the ability to detect human emotions and to develop automated emotional intelligence systems. The computer's ability to detect human emotions is gaining popularity in creating sensitive systems such as learning environments, health care systems and real-world. Improving people's health has been the subject of much research. This paper describes the formation as conceptual evidence of emotional acquisition and control in intelligent health settings. The authors of this paper aim for an unconventional approach with a friendly look to get emotional scenarios from the system to establish a functional, non-intrusive and emotionally-sensitive environment where users can do their normal activities naturally and see the program only when pleasant mood activating services are received. The context-sensitive system interacts with users to detect and differentiate emotions through facial expressions or speech recognition, to make music recommendations and mood color treatments with the services installed on their IoT devices.

\end{abstract}
\vspace{2mm}


\section{Introduction}
Emotional monitoring is important as it contains information that can help improve people's well-being. Behavioral automated modeling is essential for the development of modern applications in AI. Automatic emotional recognition has many applications in areas where equipment or machines need to communicate or monitor people. Gen-Y does not have time to visit psychiatrists as they are busy with devices connected on the internet, and during such testing times, people's mental health has taken a toll. In today's world, stigma and subjugation associated with mental health continue to plague mankind. An efficient system can greatly help people to improve their emotional state by monitoring their behavior.

To fix this subjugation; electrodes penetrating the skin are attached to the head in an electroencephalogram (EEG) for detecting human emotions by analyzing delta, theta, alpha, beta, and gamma-band waves \cite{b1}; or hefty fees are paid to psychiatrists; mood detection using facial expressions and speech sensitivity becomes need of the hour.

An alternative approach not undertaken by the authors of this paper, heterogeneous and wearable sensors send biometric data to monitor the subject, generating large amounts of sensitive data (e.g. images)\cite{b2}. Also, the location where such systems are installed should be considered and agreed upon by users, and it is considered that the sensors are located in strategic locations that require the optimal location of the target. In line with ethical standards, monitoring a person in their private living space should be considered a sensitive issue.\cite{b3}

Concretely, a person's emotional state can be reflected in a certain image or speech. This problem is widely studied in computer view especially on two sides:

(1) facial analysis \cite{b4}, and (2) posture and gesture analysis. Face Recognition (FER)\cite{b5} can be widely used in a variety of research areas, such as psychiatric diagnosis and social / physical interactions. Speech emotion recognition (SER)\cite{b6} is still a challenging task in the field of high computation because there are no defined levels of sensitivity. The speech signal contains large amounts of information related to the emotions conveyed by a person. The speech recognition system fails miserably if powerful strategies are not developed to deal with speech sensitive emotion recognition. 

The authors of this paper propose a context-aware system consisting of user-controlled device cameras such as video sensors and microphones as sound sensors, for interpreting facial expressions from static image annotations and companion text or speech sensitive emotion recognition, aimed to create a smart, advanced AI environment for sentiment analysis. A non-intrusive approach has been demonstrated, using user devices equipped with web cameras as image sensors to perform facial recognition (FER) and microphones to perform speech emotion recognition (SER). Due to the proliferation of inexpensive sensors such as cameras and microphones, as well as embedded processors, there are unprecedented opportunities for realizing real-time applications such as real computer immersion, and gesture control, and other human-centered applications such as smart and assistive environments.

Smart environments are important for the well-being of people with disabilities as it can greatly improve their daily lives. People with mobility impairments tend to choose camera and speech connectors, as this is customized, comfortable, expensive, and does not require user-borne accessories that can draw attention to their disability. Image and speech input is processed in real-time so that users can work effectively with assistive software to improve their lifestyle. 
 
The remainder of this paper is structured as follows. In Section II, a review of the previous research pertaining to Emotion recognition and Context-Aware systems is presented and in Section III, we discuss the Emotions Module of this research which includes a detailed description of the proposed system. Music recommendation and Color therapy actuation services of the proposed system have been introduced in sections IV and V, respectively followed by the analysis of the results of the experiments in Section VI and the concluding remarks in Section VII. Future work and visualization of a truly efficient, smart, natural and healthy environment system has been discussed in Section VIII.

\vspace{2mm}
\vspace{2mm}

%\begin{figure}[htbp]
%\centerline{\includegraphics[scale=0.3]{EEG.jpg}}
%\caption{Example of a EEG figure \cite{b4}}

%\end{figure}

\section{Literature Review}

%\subsection{Maintaining the Integrity of the Specifications}
Recent research has focused on improving the quality of life of a person by designing and building sensors that are directly or indirectly connected to the human body \cite{b3}. This method, previously researched in the field of intelligent design, however intrusive, is a fusion of a number of components including physiological signals (blood volume and heart rate, respiration, EEG, and skin conductance), sensory integration, human-computer connectors, networking, and extensive computing for the purpose of multimodal recognition of emotions and actuation of response services \cite{b2}. One of the most challenging domains in multi-sensor environments\cite{b7} is the automatic analysis of multi-group conversations from sensory data.

The importance of context in emotion perception is well supported by various studies in psychology. Non-verbal communication is important, as it is substantially documented in social psychology and cognition, it opens up new opportunities for new cameras and microphone-equipped spaces\cite{b8}.  Human interactions with the computer will be natural and effective when interfaces are dealing with human emotions or stress. Previous studies have focused on the acquisition of emotion via multiple sensors but the recognition of facial expression and speech sensitive emotion is gaining importance because of its broader system. 
From a computer vision perspective, most previous attempts have focused on analyzing facial expressions and, in some cases, on body movements and gestures. Some of these methods work best in certain settings. The two most popular methods used in the literature for automated FER systems are based on geometry and appearance. Most researchers use variations of Russel's circumplex model (Fig. 1) which provides a distribution of basic emotions in a dual space in relation to valence and arousal \cite{b9}: Valence (V), which measures how pleasant emotions are, from bad to good; Arousal (A), which measures a person's level of disturbance, from inactivity / calm to anger/willingness to act; and Dominance (D) which measures the level of control a person feels about the condition, from compliance / uncontrolled to in-control / management.

\begin{figure}[htbp]
\centerline{\includegraphics[scale=0.5]{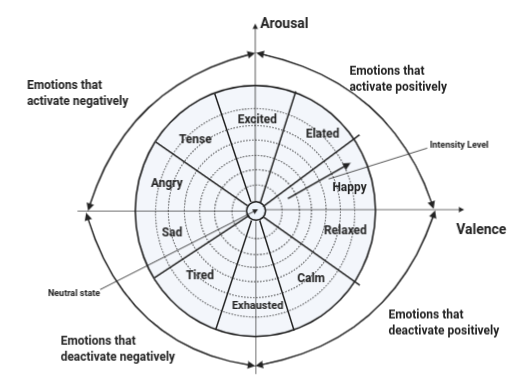}}
\caption{The Russel's circumplex model of emotions}

\end{figure}

As emotion recognition presumes the modeling of the dynamics of acoustic or visual features, some classification strategies in the field of AER make use of dynamic classifiers like Hidden Markov Models (HMM) and Dynamic Bayesian Networks (DBN) \cite{b10}. Alternative strategies apply static techniques such as Support Vector Machines (SVM) that process statistical functions of low-level features which are computed over longer data segments \cite{b11}. Quality of the human-computer interface that mimics human speech emotions relies heavily on the types of features used and also on the classifier employed for recognition \cite{b12}. A unimodal framework for short-term context modeling in dyadic interactions was proposed in \cite{b13}. 

The emotional state of a speaker can be identified from the facial expression (Ekman, 1973; Davis and College, 1975; Scherer and Ekman, 1984) \cite{b1} and spoken words (McGilloway et al., 2000; Dellaert et al., 1996; Nicholson et al., 1999). Ultimately, a combined analysis of these features leads to high accuracy of recognition. In this paper, the focus is multimodal: on facial expression and speech emotion recognition, allowing for the definition of a desired emotion and evaluating its intensity. \cite{b14}

\section{Emotions Module:}

\subsection{Dataset Construction}

Over the past decade, much effort has been put into building facial expressions and speech emotion recognizers. Because most researchers use limited data sets, the generalizability of these various methods remains unknown. Currently, technological approaches to image-related activities such as image classification and object acquisition are all based on Convolution Neural Networks (CNNs). These operations require the construction of CNN with millions of parameters. Commonly used CNN extraction features include a set of layers that are fully integrated at the end. Fully connected layers most of the parameters in a CNN \cite{b15}.

The basis of FER is the Facial Action Coding System, which incorporates facial expressions using a set of specific local facial movements, called Action Units. These facial-based approaches often use facial geometry-based features to describe the face. Afterward, the extracted features are used to identify Action Units which include regions such as eyes, cheeks, lips that are used to perceive basic emotions: anger, joy, sadness and neutral speech.

Two datasets were used in this research work. The first dataset was Extended Yale Face database (Fig. 2) consisting of:

%\begin{figure}[htbp]
%\centerline{\includegraphics[scale=0.1]{PhotoGrid_1607746189619.png}}
%\caption{Various emotions used for training and testing data\cite{b16}}

%\end{figure}

Six basic emotion classes: anger, happy, neutral, sad and disgust containing 16128 images of 28 human subjects under 9 poses and 64 illumination conditions. The dataset creator had cropped the images to face-only, performed operations like gray-scaling and resizing to 48x48. 
The second dataset was the Kaggle Facial Emotion Recognition (FER2013) dataset \cite{b16}\cite{b17} comprising of 35,887 images, each 48 x 48 pixels (8-bit grayscale) with disgust being the only underrepresented one within the Kaggle dataset (Fig. 2), at 1.

We found these datasets to be representative because of their
size, and unstructured nature of faces (in terms of facial
orientation, ethnicity, age, and gender of the subjects).
Data Preprocessing Steps:
\begin{itemize}
\item Manually cleaning the datasets to remove duplicates,
misclassified expressions and bias due to
underrepresented class labels of disgust and fear
emotions
\item Splitting the data into train, validation and test
(80:10:10 = 28275 train, 3530 test, 3532 validation were
the number of images taken)
\item Applying image augmentation using image data generator
in Tensorflow, images were normalized to 0-255, +/- 15°
rotation, +15° zoom and horizontally flipped

\begin{figure}[htbp]
\rightline{\includegraphics[scale=0.37]{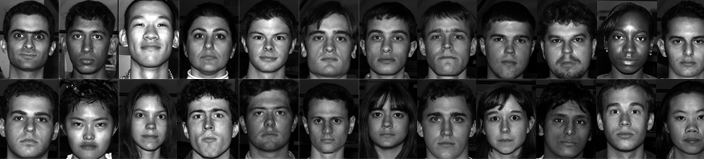}}
\caption{Dataset used for training and testing \cite{b17}}

\end{figure}

\item Haar Cascades to crop out only faces from the images
from live feed while getting real-time predictions
\end{itemize}

We chose RAVDESS dataset\cite{b18} for Speech Emotion Recognitions :-
\begin{itemize}
\item It contains speech and song files by 247 untrained Americans in eight different emotions (Calmness, joy, sadness, anger, fear, disgust, and surprise, and neutrality) at two levels of intensity. 
\item The talk file contains 1440 files: 60 attempts per actor x 24 players = 1440
\item The song file contains 1012 files: 44 trials per actor x 23 actors = 1012
\item The files are in the WAV raw audio file format and all have 16-bit bitrate and a sample rate of 48 kHz. All files are uncompressed, lossless audio, which means that audio files in the database have not lost any information/data or modified from the original recording.
\item We found this dataset to be representative because the dataset is gender-balanced with 24 professional actors, 12 males and 12 females.
\item Audio files are created in a controlled environment and each contains the same statements spoken in the American manner. Speech emotion experiments are widely done in Western side of the world so the data is usually biased towards the American accent.

\end{itemize}

Data Processing and SER Exploration essentially uses the following for feature extraction: -
\begin{enumerate}
\item Mel Scale - deals with the perception of frequency, the scale of the pitches judged by the audience to be equidistant from each other.
\item Pitch - determines how high or low the sound is. Depending on the frequency, the higher the pitch the higher the frequency.
\item Frequency - the vibration speed of sound, measuring wave cycles per second.
\item Chroma - A sound representation where the spectrum is displayed in 12 bins representing 12 different semitones (or chroma). It is calculated by summarizing the size of the log frequency spectrum across octaves.
\item Fourier Transforms - used to convert from time domain to frequency domain. Time-domain shows how the signal changes over time. The frequency-domain indicates how much signal is inside each frequency band, given a range of frequencies.\cite{b19}
\end{enumerate}
Mel-frequency cepstrum coefficient (MFCC) is the most used representation of the spectral property of voice signals. These are the best for speech recognition as it takes human perception sensitivity with respect to frequencies into consideration. For each frame, the Fourier transform and the energy spectrum were estimated and mapped into the Mel-frequency scale.
\vspace{2mm}
\vspace{2mm}

\begin{figure}[htbp]
\centerline{\includegraphics[scale=0.3]{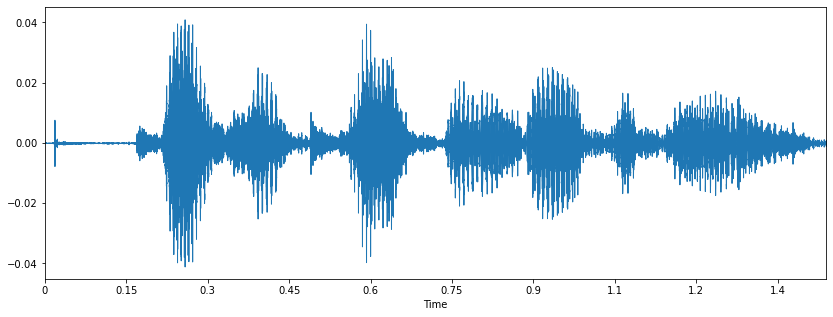}}
\caption{Image output of the audio by libROSA python library}

\end{figure}

SER Dataset preprocessing steps:

The raw audio file (.wav) is first supposed to be preprocessed in order to be classified as multi-class classification. For that, we have used Librosa library to convert the raw audio .wav file to a spectrogram image and for feature extraction like chroma, pitch, and frequency. The image thus generated is the spectrogram image and is ready to be fed in our neural network.

\begin{figure}[htbp]
\centerline{\includegraphics[scale=0.3]{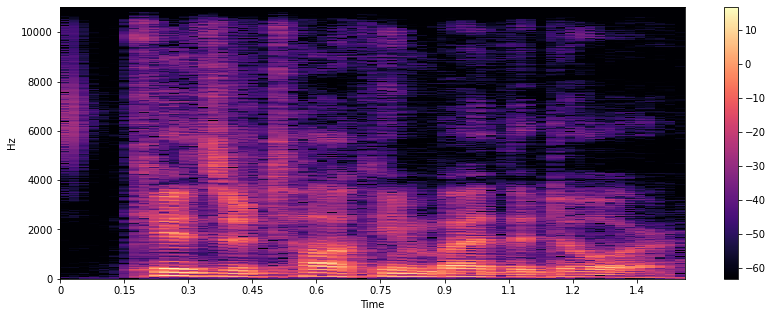}}
\caption{Mel spectrogram for a characteristic emotion}

\end{figure}

\subsection{Proposed Architecture}
The FER and SER models (Fig. 5) have been designed with the idea of attaining the best generalization accuracy. The former layer of each block represent the layers of the proposed FER model. The latter layer of each block represent the layers of the proposed SER model architecture.

We propose a model for four-class facial emotion classification (anger, happy, sad and neutral) which we evaluate in accordance to their test and generalization accuracy. These primary emotions are often referred to as “archetypal” emotions. Although these archetypal emotions cover a  rather  small  part  of  emotions in psychological research, they represent the popularly known emotions and are recommended for testing the capabilities of an automatic emotion recognizer. The FER model(Fig. 5) was designed with the idea of attaining the best generalization accuracy.

\begin{enumerate}
\item Input Layer: The input layer has fixed and predetermined dimensions. So, for pre-processing the detected face, we used OpenCV library for face detection in the image before feeding it into the layer by passing through pre-trained filters from Haar Cascades.
\item Image data generator: 48x48 size gray-scale train and validation images dataset was divided into a batch size of 64.
\item Warming-up Model: VGG-16 was used as  the  transfer learning model. After importing it, we set layers.trainable as False, and select a favorable output layer, in this case, ‘block5 conv1’. This freezes the transfer learning model so that we can pre-train or ‘warm up’ the layers of the sequential model on the given data before starting the actual training. This helps the sequential model to adjust weights by training on a lower learning rate.
\item Fine tuning model: Defining the Model, Using Sequential, the
layers in the model are as follows:

%\begin{figure}[htbp]
%\centerline{\includegraphics[scale=0.3]{cnn_model.png}}
%\caption{Architecture of FER model}

%\end{figure}

\begin{itemize}
\item GlobalAverage- Pooling2D
\item Flatten
\item Dense (256, activation: ‘relu’)
\item Dropout (0.4)
\item Dense (128, activation: ‘relu’)
\item Dropout (0.2)
\item Dense (4, activation: ‘softmax’) 
\end{itemize}

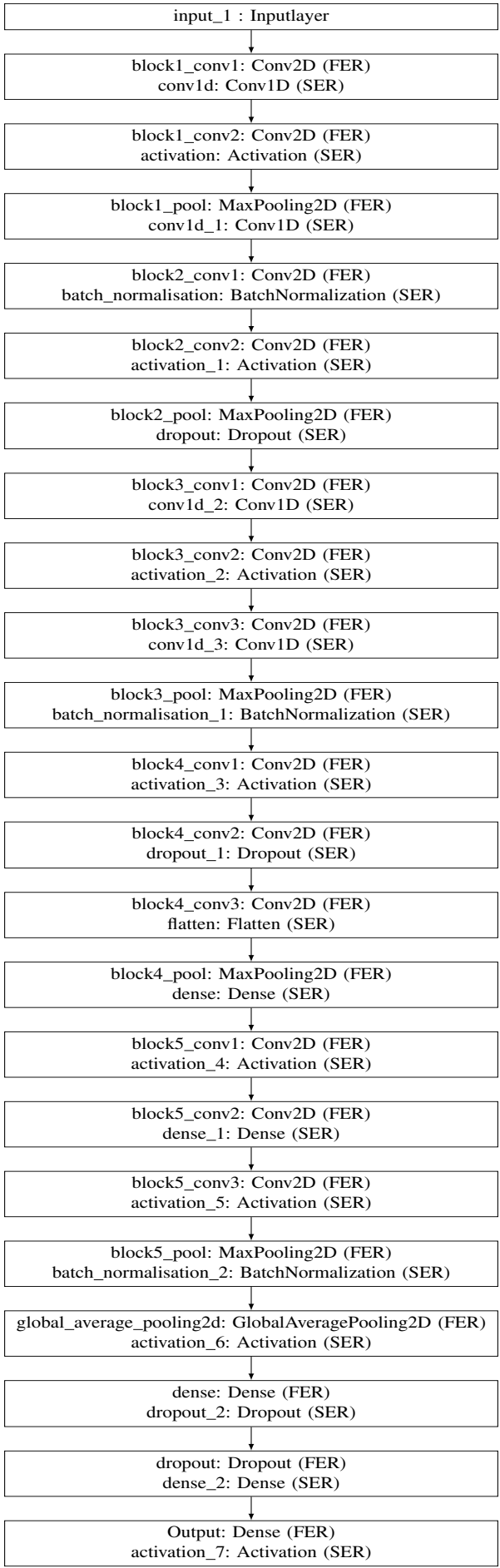
\begin{figure}
\centering

\resizebox{80mm}{250mm}{%    
    \begin{tikzpicture}[
node distance = 5mm,
  start chain = A going below,
 block/.style = {rectangle, draw, sharp corners,
                 text width=25em, align=center,
                 on chain=A, join=by -latex
                },
                        ]
\node[block]    { input\_1 : Inputlayer
\\
};% this works for one line title
\node[block]    { block1\_conv1: Conv2D (FER)\\% this works for one line title
                 conv1d: Conv1D (SER)\\
}; 
\node[block]    { block1\_conv2: Conv2D (FER)\\% this works for one line title
                 activation: Activation (SER)\\
}; 
\node[block]    { block1\_pool: MaxPooling2D (FER)\\% this works for one line title
                 conv1d\_1: Conv1D (SER)\\
};
\node[block]    { block2\_conv1: Conv2D (FER)\\% this works for one line title
                 batch\_normalisation: BatchNormalization (SER)\\
};
\node[block]    { block2\_conv2: Conv2D (FER)\\% this works for one line title
                 activation\_1: Activation (SER)\\
};
\node[block]    { block2\_pool: MaxPooling2D (FER)\\% this works for one line title
                 dropout: Dropout (SER)\\
};
\node[block]    { block3\_conv1: Conv2D (FER)\\% this works for one line title
                 conv1d\_2: Conv1D (SER)\\
};
\node[block]    { block3\_conv2: Conv2D (FER)\\% this works for one line title
                 activation\_2: Activation (SER)\\
};
\node[block]    { block3\_conv3: Conv2D (FER)\\% this works for one line title
                 conv1d\_3: Conv1D (SER)\\
};
\node[block]    { block3\_pool: MaxPooling2D (FER)\\% this works for one line title
                 batch\_normalisation\_1: BatchNormalization (SER)\\
};
\node[block]    { block4\_conv1: Conv2D (FER)\\% this works for one line title
                 activation\_3: Activation (SER)\\
};
\node[block]    { block4\_conv2: Conv2D (FER)\\% this works for one line title
                 dropout\_1: Dropout (SER)\\
};
\node[block]    { block4\_conv3: Conv2D (FER)\\% this works for one line title
                 flatten: Flatten (SER)\\
};
\node[block]    { block4\_pool: MaxPooling2D (FER)\\% this works for one line title
                 dense: Dense (SER)\\
};
\node[block]    { block5\_conv1: Conv2D (FER)\\% this works for one line title
                 activation\_4: Activation (SER)\\
};
\node[block]    { block5\_conv2: Conv2D (FER)\\% this works for one line title
                 dense\_1: Dense (SER)\\
};
\node[block]    { block5\_conv3: Conv2D (FER)\\% this works for one line title
                 activation\_5: Activation (SER)\\
};
\node[block]    { block5\_pool: MaxPooling2D (FER)\\% this works for one line title
                  batch\_normalisation\_2: BatchNormalization (SER)\\
};
\node[block]    { global\_average\_pooling2d: GlobalAveragePooling2D (FER)\\% this works for one line title
                  activation\_6: Activation (SER)\\
};
\node[block]    { dense: Dense (FER)\\% this works for one line title
                  dropout\_2: Dropout (SER)\\
};
\node[block]    { dropout: Dropout (FER)\\% this works for one line title
                  dense\_2: Dense (SER)\\
};
\node[block]    { Output: Dense (FER)\\% this works for one line title
                  activation\_7: Activation (SER)\\
};

    \end{tikzpicture}
}    
\caption{Architecture of FER and SER model}

\end{figure}

\item Training Specifications: The pre-training is done by using RMSProp at learning rate: 1e-5 and for 30 epochs. After pre-training, we set layers.trainable as True for the whole model. Now the actual training will start. It is done by taking Adam optimizer at learning rate: 1e-4 for 25 epochs. Setting the Hyper Parameters and constants (Only the best parameters are displayed below):
\begin{itemize}
\item Batch size: 64
\item Image Size: 48 x 48 x 3
\item Optimizers: o RMS Prop (Pre-Train) o Adam
\item Learning Rate: o Lr1 = 1e-5 (Pre-Train) o Lr2 = 1e-4
\item Epochs 1 = 30 (Pre-Train) o Epochs 2 = 25
\item Categorical Cross Entropy $H_(p,q)$  was chosen as loss function and the equation is given as  follows,
\[H(p,q)\ =\ -\sum _{i=1} ^ {n} p_{i}\log q_{i}\    \  \  \   \  (1)\] 
where  $ p_{i}$ is the truth label and $q_{i}$ is the softmax probability for $i^{ith}$ class and $n$ is the number of classes or 4.
\end{itemize}
\end{enumerate}

We propose a model(Fig. 5) for sixteen-class Speech emotion recognition classification (neutral, calm, happy, sad, angry, fearful, disgust, surprise, 8 for female and 8 for male speech) which we evaluate in accordance with their test and generalization accuracy. The model was designed to get the best generalization accuracy.

\begin{enumerate}
\item Image Data Segregation: All the images are of RGB Channels, separated with a fixed batch size of 16, and are segregated based on the actual 16 emotion classes.

\item A custom class of neural networks has been defined to achieve the goal. In this class, we first defined the core sequential model with the image data generator as a pre-training parameter. Convolution 1-D Layers along with normal Dense layers are used for the core model. Since this was a multi-class classification, we implemented a Conv1D Model along with a ’selu’ activation function initially and a ‘softmax’ activation function in the end with a relevant optimizer like ‘adam’ to implement the core model for this purpose.

\item Fine Tuning: After Defining the initial architecture, the following were implemented to ensure the smooth process of training and to expect a better-generalized accuracy:
\begin{itemize}
\item Conv1D(256,input shape=(66,1))
\item Activation(‘selu’)
\item Batch Normalisation
\item Dropout(0.4)
\item Flatten
\item Dense
\item Activation(‘selu’)
\item Batch Normalisation
\item Dropout(0.4)
\item Dense(16)
\end{itemize}

\item Training Specifications: The training has been ensured to have a smooth learning process with help of ‘Adam’ Optimiser with learning rate=0.001 for 65 epochs. Early callbacks have been imposed with a patience=10, to wait initially for achieving better accuracy and minimizing the loss function along with stopping the training if the learning worsens. Also, we defined a custom class to directly export the model with the best results achieved,i.e, best accuracy in the entire training epochs in between, and to keep track of whether the next epoch had a better accuracy or the previous one.

\end{enumerate}

%\begin{figure}[htbp]
%\centerline{\includegraphics[scale=0.32]{SER.png}}
%\caption{Architecture of SER model}

%\end{figure}

\subsection{Emotion Recognition Module}

The system has to decide automatically the proper reaction in response to the detected emotion, in order to simulate a positive response in the user. Decision making can vary from one user to another and also in accordance with his/her health evolution. It is in “Decision making” where the intelligence of the system lies \cite{b20}. Considering the emotion detected the system reacts and decides the action to execute. For example, considering that our system reacts to the user’s mood, once this is detected, the system will vary the color and change the music. For instance, if the     action is to create a relaxing environment, the system will launch commands addressed to project warm colors, and play pleasant music (in broad terms) \cite{b21}.A variety of application services in ambient intelligence environments can be realized at reduced cost by encapsulating attractive services like music recommendation \cite{b22} and color/light-actuation in middleware applications that are shared by the IoT connected devices. High level abstractions offered by middle-ware infrastructures make it possible to hide the complexities in ambient intelligence environments.

\begin{figure}[h!]
\centerline{\includegraphics[scale=0.14]{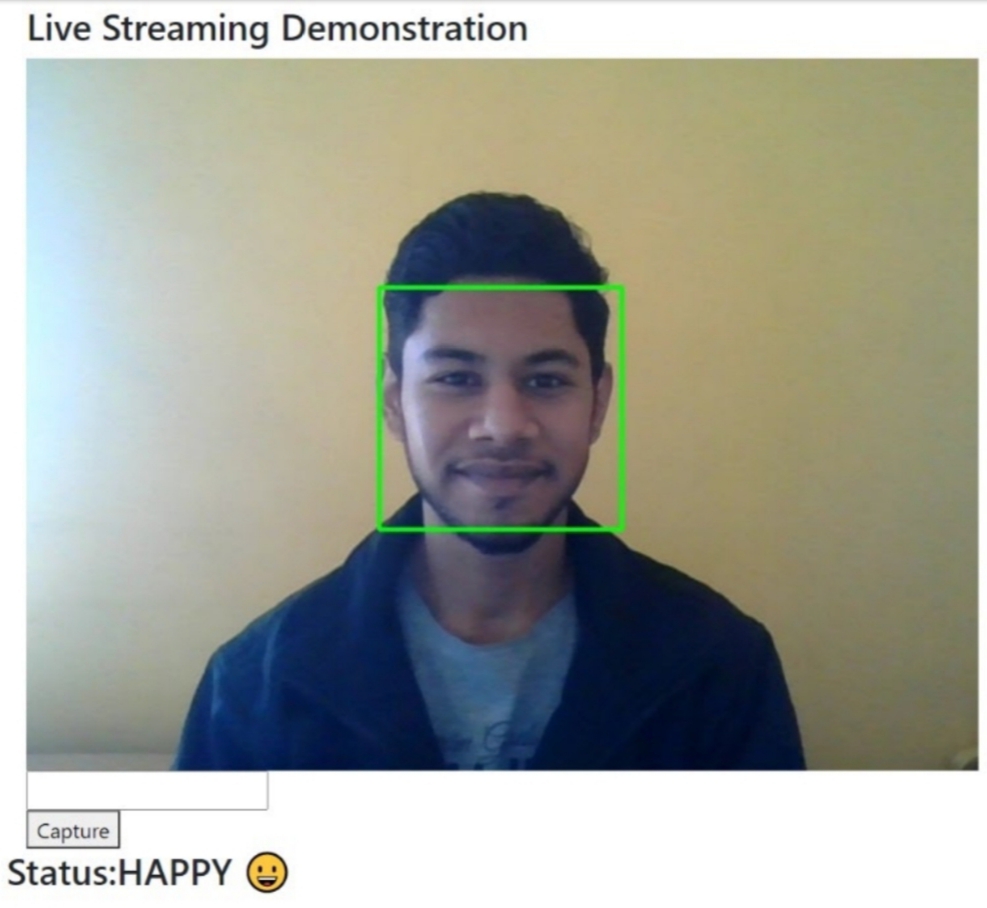}}
\caption{A glimpse of our application}
\end{figure}
\section{Music Recommendation Module }

Balkwill and Thompson found that even listeners who are unfamiliar with the tonal system are sensitive to the emotion expressed by a piece of music. Hunter et al. explored how tempo and musical harmony affect the emotional responses of listeners \cite{b23}. They found that quick-tempo and high-speed routes add to the excitement, while slow-moving, low-key modes make listeners feel sad. (Interestingly, both levels of happiness and sorrow were high when fast tempo was combined with major modes or slower tempos with minor modes.) These findings are consistent with Hevner's conclusions, and support the use of tempo and tonality modulation within the proposed application. Zentner, Grandjean, and Scherer's Lundqvist et al.\cite{b7} demonstrated that listeners' responses to happy stimuli generated “more zygomatic facial muscle activity, more pleasure, and less sadness” than sad stimuli which is often seen in greater magnitude than it may sound. These results can be used as a basis for building a response to the proposed plan. Of Saarikallio's seven strategies for MMR, AMAI \cite{b23} focuses on discharge ("venting anger or sadness through music that expresses these emotions"), diversion (''forgetting thoughts and feelings with the help of good music "), and progression from discharge to diversion. Given the results of Lundqvist et al, their findings have confirmed that the use of subsequent diversion can have a consistent effect on all listeners rather than the use of diversion technique alone.

In this paper we described the design, and implementation of a system for generation and playback of music that adapts to listeners’ affective states with the goal of increasing positive affect- “diversion”. In particular, we explored the effectiveness of employing the diversion MMR strategy. A case has been made of dynamic music that has the ability to direct the listener results more efficiently 
\begin{figure}[htbp]
\centerline{\includegraphics[height=6.2cm,width=9.3cm]{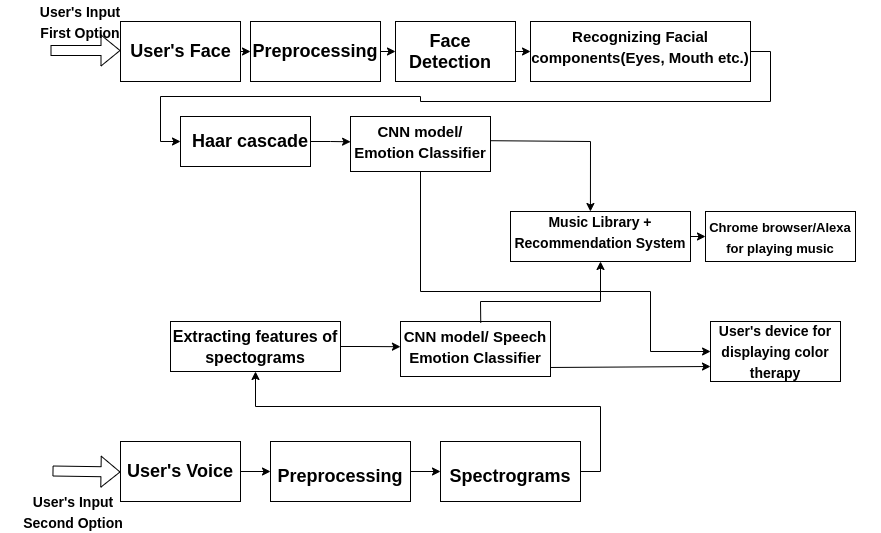}}
\caption{Architecture of proposed system}

\end{figure}
than standard music such as MP3, like in the case of AMAI, the conversion from discharge to diversion was controlled by the user's situation, with the system having control of the precise timing of when to switch sections and strategies.
Using Selenium automation in Python, whenever you make prediction from the proposed model, you get a word as emotion – ‘ANGRY’, ‘HAPPY’, ‘NEUTRAL’, ‘SAD’ which is used in automation\cite{b16} for parsing the YouTube web pages using ‘chromedriver 87.0.4280.88’ for automatically playing music to simulate a positive effect when users’ facial emotion is detected, and redirects to the recommended YouTube MP4 video.

\section{Color/Light Actuation}

Color is another factor that is omnipresent in an environment. Moreover, evidence of the influence of color on human emotions is validated by metonymic and metaphoric thinking, formation of specific emotional reactions for color perception, and sharing connotative structure in the language for color and emotion terms. Color is a powerful communication tool and can be used to express action, influence emotions, and influence the body's response. Certain colors are associated with increased blood pressure, weight gain, and eyestrain. Many authors emphasize that certain colors have a profound effect on mood and control. For instance, in a trial, muscle strength is decreased within 2: 7 s inmates in prisons who resided for a limited time in bubble-gum pink cells (Baker-Miller pink) cells \cite{b7}. It is important to note that it has been shown that color features such as chroma, hue, or light also have an effect on emotions. It is known that while perceiving
color, the brain associates it with a particular emotion. This phenomenon is known as color emotion. One can appreciate the colors that provide the desired change from negative to positive according to previous works. Also, depending on the initial sensitivity to the orthogonal scale, we have(fig. 8):

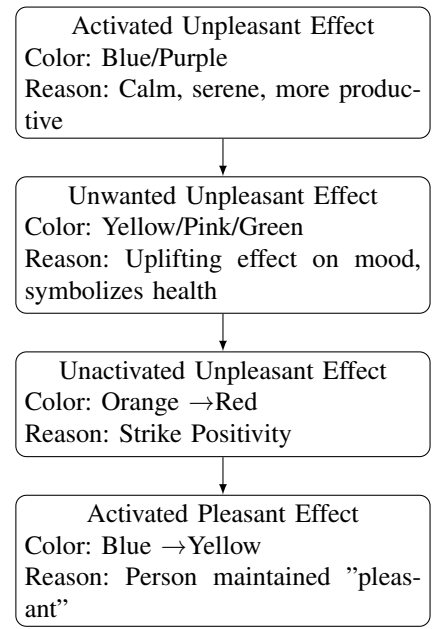
\begin{figure}[h!]
    \centering
    \begin{tikzpicture}[
node distance = 5mm,
  start chain = A going below,
 block/.style = {rectangle, draw, rounded corners,
                 text width=15em, align=justify,
                 on chain=A, join=by -latex
                },
                        ]
\node[block]    {\hfil  Activated Unpleasant Effect\\% this works for one line title
                 Color: Blue/Purple\\
                 Reason: Calm, serene, more productive};
\node[block]    {\hfil  Unwanted Unpleasant Effect\\% this works for one line title
                 Color: Yellow/Pink/Green\\
                 Reason: Uplifting effect on mood, symbolizes health}; 
\node[block]    {\hfil  Unactivated Unpleasant Effect\\% this works for one line title
                 Color: Orange \textrightarrow Red\\
                 Reason: Strike Positivity}; 
\node[block]    {\hfil  Activated Pleasant Effect\\% this works for one line title
                 Color: Blue \textrightarrow Yellow\\
                 Reason: Person maintained "pleasant"};                   
    \end{tikzpicture}

\caption{Schematic flowchart of Color/Light Actuation}
\vspace{2mm}
\end{figure}
\begin{figure}[htbp]
\centering
\includegraphics[height=2.6cm,width=8.3cm]{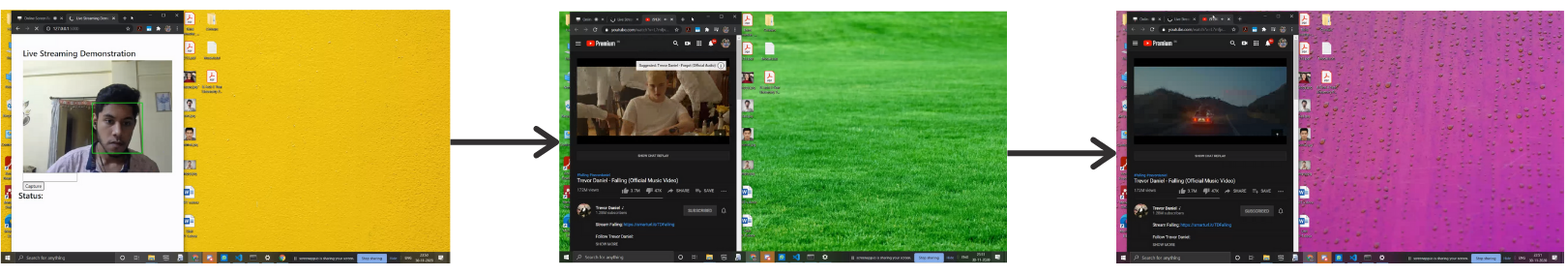}
\caption{Application of the color therapy for emotion sad}

\end{figure}

\section{Results}

In facial emotion recognition, we were able to achieve a four-class generalization accuracy of 77.92\%, test  accuracy  75.39\%,  and  train  accuracy of 87.14\% for classifying an image of a face under labels “happy”, “sad”, “angry”, “neutral”. In speech sensitive emotion recognition, we were able to achieve a validation accuracy of 77.3\%, test accuracy of 73.3\% and train accuracy of 93\% for classifying 8 different emotions in speech.

With the emerging advanced technologies in hardware and sensors, FER and SER systems \cite{b24} have been developed to support real-world application scenes, instead of laboratory environments \cite{b25}. Although the laboratory-controlled systems achieve very high accuracy, around 97\%, the technical trans- ferring from the laboratory to real-world applications faces a great barrier of very low accuracy, approximately 50\%.

\begin{figure}[htbp]
\centerline{\includegraphics[scale=0.5]{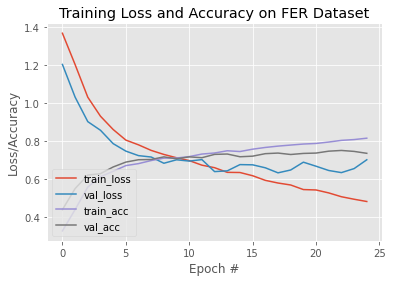}}
\caption{Training Loss and Accuracy of FER model}

\end{figure}

\begin{figure}[htbp]
\leftline{\includegraphics[scale=0.2]{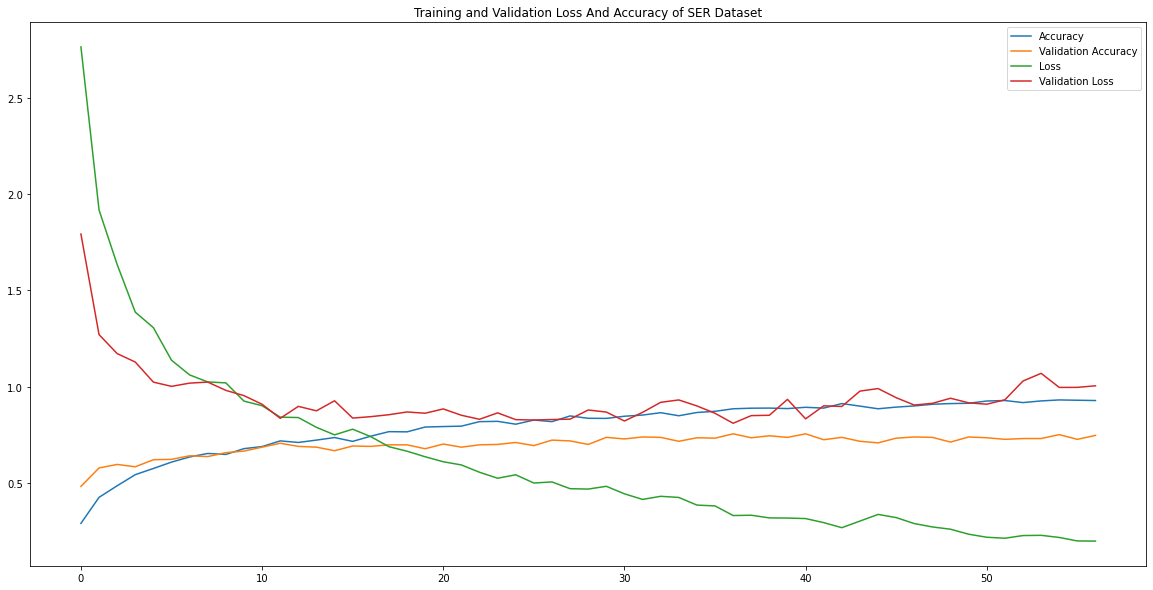}}
\caption{Training Accuracy of SER model}

\end{figure}

%\begin{figure}[htbp]
%\rightline{\includegraphics[scale=0.2]{loss.png}}
%\caption{Training loss of SER model}

%\end{figure}

Research in the field of color therapy has shown that changes in the color of our environment cause a clear emotional response. Color preferences have been shown to be closely related to personality traits and emotions. Color tone is managed in areas of the brain that deal with emotions and feelings. 

Undoubtedly, user feedback related to color depends on a large set of external factors, such as gender, age, culture, preferences, emotions, and content (e.g. time of day or location). Identifying emotional state and interpreting well using deep-learning algorithms \cite{b26}\cite{b27} has proven to be complex due to the high variability of samples in each activity.

\begin{table}[htbp]
\footnotesize
\caption{Train, Test, and Generalization Accuracy after 30 epochs(FER) and 80 epochs(SER}
\begin{center}
\begin{tabular}{|c|c|c|c|}
\hline
\cline{2-4} 
\textbf{} & \textbf{\textit{Train acc. (\%)}}& \textbf{\textit{Test acc. (\%)}}& \textbf{\textit{Generalization Acc. (\%)}} \\
\hline
FER Model& 87.14\%& 75.39\%& 75.39\% \\
\hline
SER Model& 93\%& 73.3\%& 77.3\% \\
\hline
\end{tabular}
\end{center}
\end{table}

\section{Conclusion}

The main purpose is to keep the user's emotional state healthy. The proposed constructs for the acquisition of emotions and their regulation in the Smart Home Environment are novel, open, and flexible. As a result, the detection of facial expressions and speech is done in a non-intrusive manner. The architecture is not pre-designed to target a particular health care problem, but should be able to deal with most emotionally-related problems. Mixing multimodal data significantly increased recognition levels compared to unimodal programs: multimodal method demonstrated more than 10\% improvement in relation to the most successful unimodal system. Moreover, decision making of the smart context-aware system environment improved. This paper described the complete structure (fig. 7) of providing all the necessary functions and interfaces for emotional recognition and control. First, the purpose is to determine the user's feelings by analyzing facial expressions or sensitive speech emotions. The algorithm has three stages. In the image processing phase, the facial region and facial action elements are extracted. Haar Cascade is adopted to extract the facial region from an image. Speech sensitive emotion recognition was performed in two main steps: feature (mel scale, pitch, frequency, chroma, fourier transform) extraction and classification by employing deep-learning algorithms. Then, the system makes decisions to steer the emotions of the user to a positive state with music and color/light actuation. On the output end, music and color rendering is done by changing the color of their IoT devices from a variety of colors to soften the acquired mood, for instance, a person in an angry mood is toned down by colors lying on the soft side of the color spectra whereas a person feeling sad is uplifted by bright colors for driving the user to a pleasant state of mood.

\section{Future Work}

Ingenuine and forced expressions pose a challenge to actual emotion detection and probing systems towards a healthier and happier environment  for  the  user.  A  system  operating in the real world would not rely completely on single or separate information sources due to problems such as noise, false positives or occlusions.

The difficulty in the introduction of a new intelligent system is in the uncertainty of its effect in the society. The benefit of any system appears when it is implemented deeply into the society and impacts social activities in a large scope. Social simulation, especially, through a method of inducing mood, will provide an important tool to evaluate how the system may change the behavior of the society. Researchers will be in permanent contact with the users to solve problems that may arise and to monitor the level of the users’ satisfaction,   so as to improve the system’s feedback in regulating emotions by stimuli such as color and music.

Colors can be subjective - what can make one person feel happy can make another person feel annoyed depending on the user's past experience or cultural differences. Therefore, the smart context-aware environment system should be tailored to each user according to his or her preference. Providing immersion in AR, and giving listeners an active agency (rather than a system that controls all aspects of music) can further improve the system's performance. Therefore, some user-associated information is welcome.

While recent researches show much promise, they are first and foremost indicative of the fact that there’s a long way to go before we arrive at true user-centered approach to ambient intelligence systems, visual lie-detectors and threat detection security systems, which can combine automated facial emotion, body language analysis and speech sensitive emotion for spotting potentially risky situations. Relevant to the development of AI systems that can make an environment truly intelligent is the need for such an environment to be able to:

\begin{enumerate}
    \item learn habits, preferences and needs of the specific user,
    \item enhance user capabilities and comfort by providing new and automated service execution,
    \item enhance user capabilities and comfort by
providing new and automated service execution, 
    \item integrate IoT connected elements like smart color blinds, smart home and mobile robots, and
    \item provide a structured way to analyze, decide and react over that smart, friendly environment.
\end{enumerate}

\end{document}